\documentclass[a4paper,onecolumn,11pt,accepted=2022-09-01]{quantumarticle}
\pdfoutput=1
\usepackage[utf8]{inputenc}
\usepackage[english]{babel}
\usepackage[T1]{fontenc}
\usepackage{amsmath,amssymb,amsthm,amstext}
\usepackage[colorlinks = false,hidelinks]{hyperref}
\usepackage[sort&compress,numbers]{natbib}
\usepackage{tikz}
\usepackage{lipsum}

\newcommand{\tr}[1]{\mathrm{tr }{#1}}

\newcommand{\del}[0]{\partial}

\newcommand{\swp}{\text{\texttt{SWAP}}}

\usepackage{scalerel,stackengine}
\stackMath
\newcommand\reallywidehat[1]{%
\savestack{\tmpbox}{\stretchto{%
  \scaleto{%
    \scalerel*[\widthof{\ensuremath{#1}}]{\kern-.6pt\bigwedge\kern-.6pt}%
    {\rule[-\textheight/2]{1ex}{\textheight}}
  }{\textheight}%
}{0.5ex}}%
\stackon[1pt]{#1}{\tmpbox}%
}

\newcommand{\ket}[1]{|#1\rangle}               
\newcommand{\bra}[1]{\langle #1|}              

\begin{document}

\title{Ancilla-free continuous-variable SWAP test}

\author{T.J.\,Volkoff}
\affiliation{Theoretical Division, Los Alamos National Laboratory, Los Alamos, NM 87545, USA.}
\author{Yi\u{g}it Suba\c{s}{\i}}
\affiliation{Computer, Computational, and Statistical Sciences Division, Los Alamos National Laboratory, Los Alamos, NM 87545 USA.}

\begin{abstract}
We propose a continuous-variable (CV) SWAP test that requires no ancilla register, thereby generalizing the ancilla-free SWAP test for qubits. In this ancilla-free CV SWAP test, the computational basis measurement is replaced by photon number-resolving measurement, and we calculate an upper bound on the error of the overlap estimate obtained from a finite Fock cutoff in the detector. As an example, we show that estimation of the overlap of pure, centered, single-mode Gaussian states of energy $E$ and squeezed in opposite quadratures can be obtained to error $\epsilon$ using photon statistics below a Fock basis cutoff $O(E\ln \epsilon^{-1})$. This cutoff is greatly reduced to $E + O(\sqrt{E}\ln \epsilon^{-1})$ when the states have rapidly decaying Fock tails, such as coherent states. We show how the ancilla-free CV SWAP test can be extended to many modes and applied to quantum algorithms such as variational compiling and entanglement spectroscopy in the CV setting. For the latter we also provide a new algorithm which does not have an analog in qubit systems. The ancilla-free CV SWAP test is implemented on Xanadu's 8-mode photonic processor in order to estimate the vacuum probability of a two-mode squeezed state.
\end{abstract}
\maketitle

\section{Introduction}\label{sec:intro}
Proposals for implementing quantum algorithms using photonic or continuous-variable (CV) quantum systems are motivated by the fact that CV systems are candidates for universal, fault-tolerant quantum computation \cite{PhysRevLett.97.110501,PhysRevLett.112.120504,PhysRevX.8.021054,PhysRevA.101.012316,Bourassa2021blueprintscalable}. Quantum algorithms such as Deutsch-Jozsa \cite{dj1,dj2} and Grover search \cite{grov} have counterparts in CV systems that make use of a linear optical implementation of the quantum Fourier transform (which is just a $\pi/2$ phase shift), and finite-precision homodyne measurements. 

The SWAP test is a basic quantum algorithm for estimation of the pure state overlap  $\vert \langle \psi \vert \phi\rangle\vert^{2}$ ($\tr(\rho \sigma)$ for mixed states $\sigma$ and $\rho$) originally introduced in the context of quantum fingerprinting \cite{PhysRevLett.87.167902}. In its original formulation, the algorithm requires 1. a tripartite register $ABC$, 2. a Hadamard gate on $C$, followed by a controlled-\texttt{SWAP} gate acting on $AB$ and controlled on $C$ that maps $\ket{\psi}_{A}\ket{\phi}_{B} \mapsto \ket{\phi}_{A}\ket{\psi}_{B}$ conditional on $\ket{1}_{C}$, followed by a final Hadamard gate on $C$, 3. computational basis measurement of ``ancilla'' register $C$. The quantum circuit is shown in Fig.~\ref{fig:aaa}a.~\footnote{Alternatively, it can be considered an algorithm for the BQP promise problem that requires to decide between $\vert \langle \psi \vert \phi \rangle\vert\le \delta $ (YES) or $\vert \langle \psi \vert \phi \rangle\vert=1$ (NO), given that one of these cases holds.} 
The main drawback of this algorithm is the use of the controlled-SWAP gate, which is both costly in terms the number of gates and precludes the possibility of parallelization. This is the motivation for the ancilla-free algorithms discussed in this paper (which are also, therefore, free of controlled unitaries).

When applied to quantum states of single photons, the success probability of the SWAP test is seen to be equal to the coincidence fraction in the Hong-Ou-Mandel effect \cite{PhysRevA.87.052330}. This identification motivated the discovery of an ancilla-free, destructive SWAP test for qubits with no ancillas. It was later shown that this version of the SWAP test can be identified by an algorithm that optimizes quantum circuits with respect to depth, in this case depth two \cite{cincio}. 
Applications of the SWAP test and its multi-register generalizations include variational quantum algorithms for estimation of rank, quantum entropies, and quantum Fisher information \cite{PhysRevResearch.3.033251}, entanglement spectroscopy and estimation of polynomial functions of quantum states \cite{PhysRevB.96.195136,entspec,entspec2,brun}, virtual cooling and error mitigation \cite{PhysRevX.9.031013,PhysRevX.11.041036}, and implementing nonlinear transformations of quantum states \cite{holmes2021nonlinear}.

In this work, we present a destructive, ancilla-free CV SWAP test that makes use of linear optical operations and photon number-resolving detection. The algorithm is not a generalization of the destructive, ancilla-free SWAP test for qubits, but rather utilizes the fact that eigenvectors of the CV $\swp$ gate are images of tensor products of Fock states under a 50:50 beamsplitter. We show how, given a desired precision $\epsilon$ of the overlap estimate and the energy $E$ of the input states, one can determine a  Fock basis cutoff $M(\epsilon, E)$ for the photon number-resolving measurement that is sufficient to achieve this precision. These methods are illustrated for pairs of equal-energy CV Gaussian states such as squeezed and anti-squeezed vacuum and antipodal coherent states. After providing example applications of the ancilla-free CV SWAP test, our last main result introduces the ancilla-free, CV PERM test, which allows to calculate $\text{tr}\rho^{L}$, $L\ge 2$, for any CV state $\rho$ by measurement of the cyclic permutation operator.

The structure of the paper is as follows. We conclude the introduction with a primer on multimode CV systems, where we introduce the main mathematical concepts and notation needed for the presentation of the results of this work. In Section \ref{sec:rev} we review the ancilla-free destructive SWAP test of Refs. \cite{PhysRevA.87.052330,cincio} for qubits and generalize this algorithm to qudits. In Section \ref{sec:ooo} we discuss the main result of this paper: an ancilla-free CV SWAP test algorithm. In Section \ref{sec:x8} we provide a proof-of-principle experimental implementation of this algorithm on Xanadu's X8 device by estimating the vacuum probability of two-mode squeezed states. 
Section \ref{sec:applic} is dedicated to applications. In Section \ref{sec:spec}, we  first show that cost functions utilized in variational compiling of CV quantum circuits can be computed by a generalization of the two-mode, ancilla-free CV SWAP test. As a second application, in Section \ref{sec:tct},
we give a CV generalization of the so-called ``two-copy test'' \cite{entspec} which is then used to estimate $(\text{tr}\rho^{n})^{2}$ using $2n$ copies of a purification of a generally mixed state $\rho$.
In Section \ref{sec:ext} we give a new algorithm for calculating $\tr (\rho^n)$, which directly uses $n$ copies of $\rho$. Both the two-copy test of Section \ref{sec:tct} and the algorithm of Section \ref{sec:ext} have advantages over the conventional Hadamard test for entanglement spectroscopy. 
In Section \ref{sec:dvcv} we combine the discrete-variable (DV), ancilla-free SWAP test of Section \ref{sec:rev} and the CV, ancilla-free SWAP test of Section \ref{sec:ooo} to give an ancilla-free SWAP test valid for hybrid DV-CV states. 

\subsection{Primer on multimode continuous-variable systems\label{sec:cvintro}}

Let $L^{2}(\mathbb{R})$ (the set of square integrable measurable functions on $\mathbb{R}$) have orthonormal basis given by the eigenfunctions of a quantum harmonic oscillator of unit frequency and unit mass. We define an $M$-mode continuous-variable system as $L^{2}(\mathbb{R}^{M})\cong \ell^{2}(\mathbb{C})^{\otimes M}$, where $\ell^{2}(\mathbb{C})$ is the set of square summable sequences of complex numbers. This is a representation of the canonical commutation relation algebra $[a_{i},a_{j}^{\dagger}]=\delta_{i,j}$, $i,j \in \lbrace 1,\ldots, M\rbrace$, where $a_{j}^{\dagger }={1\over \sqrt{2}}\left( x_{j}-{\del\over \del x_{j}}\right)$ \cite{bratteli2}. The vacuum  wavefunction proportional to $\Phi_{0}(x):=e^{-{1\over 2}\sum_{j=1}^{M}x_{j}^{2}}$ corresponds to the state $\ket{0}_{1}\otimes \cdots\otimes \ket{0}_{M}$. More generally, the wavefunction proportional to $\prod_{j=1}^{M}\left( x_{j}-{\del\over \del x_{j}}\right)^{n_{j}}\Phi_{0}(x)$ corresponds to the Fock state $\ket{n_{1}}_{1}\otimes \cdots \otimes \ket{n_{M}}_{M}$. 

A distinguished set of CV states are the Gaussian states, characterized by Wigner functions that are Gaussians on phase space $\mathbb{R}^{2M}$ \cite{holevo}. In this work we make use of three unitary operations that map the set of Gaussian states to itself:

\begin{align}
    D(\alpha)_{j}&= e^{\alpha a_{j}^{\dagger}-h.c.} \; \; \text{(displacement)}\nonumber \\
    S(z)_{j} &= e^{{1\over 2}(\overline{z}a_{j}^{2}-h.c.)} \; \; \text{(single mode squeezing)}\nonumber \\
    U_{\text{BS}}(\theta,\phi)_{i,j}&=  e^{\theta (e^{i\phi}a_{i}^{\dagger}a_{j} - h.c.)}\; , \; i\neq j \; \; \text{(beamsplitter)}
    \label{eqn:cvst}
\end{align}
with $\alpha , z\in \mathbb{C}$ and $\theta \in [0,\pi]$, $\phi \in [0,2\pi)$. We will also utilize phase space rotations, which take the form $\prod_{j=1}^{M}e^{-i\phi_{j}a_{j}^{\dagger}a_{j}}$. In Section \ref{sec:ooo}, we utilize two-mode squeezed states, defined by
\begin{equation}
    \ket{\text{TMSS}_{r}}_{1,2}:= U_{\text{BS}}\left({\pi\over 4},\pi\right)_{1,2}S(r)_{1}\otimes S(-r)_{2} \ket{0}_{1}\otimes \ket{0}_{2}
\end{equation}
 and in Section \ref{sec:ooo} we utilize CV coherent states, i.e., states of the form $\ket{\alpha}:= D(\alpha)\ket{0}$.

In this work, we consider CV systems to be read out by photon number-resolving measurement. This is a projective measurement with outcome indexed by $\mathbb{Z}_{\ge 0}^{\times M}$ corresponding to a $M$-mode Fock state.

\section{\label{sec:rev}Ancilla-free, destructive discrete variable (DV) SWAP test}

It is well-known that the expected value of $\swp$ (considered as an observable) in a product state on two registers is given by the trace of the product of the density operators on both registers \cite{PhysRevLett.104.157201}, i.e.
\begin{equation}
    \langle \texttt{SWAP} \rangle_{\rho\otimes \sigma} = \text{tr}\text{\texttt{SWAP}}\rho\otimes \sigma = \text{tr}\rho \sigma \, .
    \label{eqn:swpswp}
\end{equation}
It suffices to know how to implement $\swp$ between two registers since $\swp$ between multiple registers is defined as the tensor products of the former.

\begin{figure*}[t]
\centering
\includegraphics[width=0.9\textwidth]{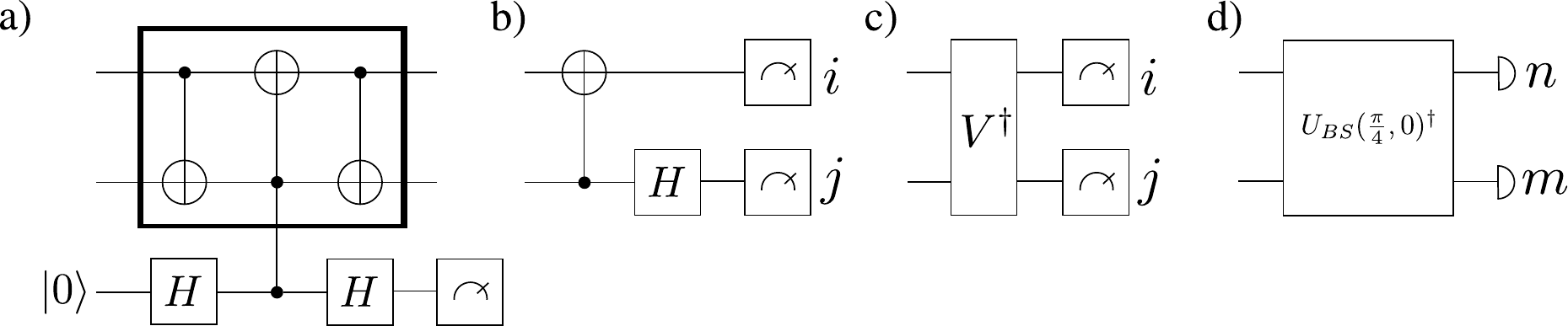}
\caption{\label{fig:aaa}(a) SWAP test circuit with ancilla. Bold box is $\swp$. b)-d) Ancilla-free destructive SWAP test circuits for states of b) qubits ($H$ is the Hadamard gate), c) general DV modes (i.e., qudits; $V$ is a unitary transformation such as in (\ref{eqn:vvv}) taking the qudit computational basis to the eigenbasis of the qudit $\swp$ operator), d) CV modes ($U_{\text{BS}}\left( {\pi\over 4},0\right)$ is the 50:50 beamsplitter defined in (\ref{eqn:cvst})).}
\end{figure*}

We first discuss the destructive, ancilla-free SWAP test algorithm for qubits for which we know an efficient quantum circuit in terms of elementary gates. For two qubits, the eigenvectors of \texttt{SWAP} are given by the Bell basis \begin{align}
    \ket{\Phi_{0,0}}&\equiv {1\over \sqrt{2}}\left( \ket{0}_{A}\ket{0}_{B} + \ket{1}_{A}\ket{1}_{B} \right) \nonumber \\
    \ket{\Phi_{0,1}}&\equiv {1\over \sqrt{2}}\left( \ket{0}_{A}\ket{0}_{B} - \ket{1}_{A}\ket{1}_{B} \right)\nonumber \\
    \ket{\Phi_{1,0}}&\equiv {1\over \sqrt{2}}\left( \ket{1}_{A}\ket{0}_{B} + \ket{0}_{A}\ket{1}_{B} \right)\nonumber \\
    \ket{\Phi_{1,1}}&\equiv {1\over \sqrt{2}}\left( \ket{1}_{A}\ket{0}_{B} - \ket{0}_{A}\ket{1}_{B} \right)
\end{align}
with respective eigenvalues $\lambda_{i,j}=(-1)^{ij}$.
A measurement in the Bell basis can be implemented by a computational basis measurement preceded by a quantum circuit $C^{\dagger}$, where $C$ satisfies $C\ket{i}_{A}\ket{j}_{B}=\ket{\Phi_{i,j}}$ such as $C=CNOT_{BA}H_{B}$ -- see Fig.~\ref{fig:aaa}b).

For $\rho$ and $\sigma$ generally mixed states of $K$ qubits, the expectation value in \eqref{eqn:swpswp} can be estimated by preparing the two states, measuring pairs of qubits in each system in the Bell basis, and repeating this $S$ times. The following then is an unbiased estimator of the overlap $\text{tr}\rho\sigma$:
\begin{align}
    \reallywidehat{\text{tr}\rho\sigma} &= \frac{1}{S} \sum_{s=1}^S \prod_{k=1}^{K} (-1)^{i_k(s) j_k(s)}
    \label{eqn:estimator1}
\end{align}
where $i_k(s),j_k(s)$ are the outcomes of the Bell measurements of the $k$-th pair of qubits in the $s$-th run of the experiment. 
The Berry-Esseen theorem implies that the difference between the distribution of the estimator and the normal distribution $\mathcal{N}(\text{tr}\rho\sigma,O(1/S))$ is at most $O(1/\sqrt{S})$ \cite{durrett}.

The appropriate generalization of the Bell basis measurement to the case of qudit registers $AB$ is the projective measurement defined by the eigenvectors of \texttt{SWAP}. These can be constructed using the fact that for a bipartite system $AB$ with $\text{dim}A=\text{dim}B=d$, $\text{\texttt{SWAP}}=P_{+}-P_{-}$ where $P_{+}$ ($P_{-}$) is the projector to the symmetric (antisymmetric) subspace of the Hilbert space $AB$. 
Since the spectrum of $\swp$ is highly degenerate there are many unitaries that diagonalize it.
One such unitary is given by
\begin{equation}
    V\ket{i}\ket{j} = \begin{cases} \ket{i}\ket{i} & j=i\\
    {1\over \sqrt{2}}\left( \ket{i}\ket{j} + \ket{j}\ket{i} \right) & i<j \\ {1\over \sqrt{2}}\left( \ket{i}\ket{j} - \ket{j}\ket{i} \right) & i>j \end{cases}
    \label{eqn:vvv}
\end{equation}
which generates the \texttt{SWAP} eigenvectors having respective eigenvalues $\lambda_{i,j}=(-1)^{1(i>j)}$ where $1(i>j)$ is the indicator function of the $i>j$ subset of $\mathbb{Z}_{d}^{\times 2}$. Therefore,  the circuit in Fig.~\ref{fig:aaa}c allows to estimate (\ref{eqn:swpswp}) for a system of qudits. 

Unitary $V$ of Eq. \eqref{eqn:vvv} is one example that is easy to specify on paper but might not be the easiest to implement in terms of a given gate set. For example, for $d=2$, $V$ does not correspond to the unitary of Fig.~\ref{fig:aaa}b and its quantum circuit has two CNOTs as opposed to one. 
To the best of our knowledge, short-depth gate decompositions of any $V^\dagger$ using a standard gate set are not known for $d>2$ and it is an open question if constant depth circuits for this task exist analogous to the qubit and CV cases. We note that, unlike qubits, for $d>2$, any orthonormal basis consisting of eigenvectors of \texttt{SWAP} does not coincide exactly with the set of qudit Bell states $\ket{\Phi_{z,x}}$ defined by
\begin{equation}
    \ket{\Phi_{z,x}}\equiv \left( X(x)\otimes Z(z) \right) {1\over \sqrt{d}}\sum_{j=1}^{d}\ket{i}_{A}\ket{i}_{B}
\end{equation} where $(z,x)\in \mathbb{Z}_{d}^{\times 2}$. In Appendix \ref{sec:app1} we show how \texttt{SWAP} eigenvectors can be formed as superpositions of pairs of qudit Bell states. However, this fact by itself does not lead to an efficient gate decomposition.

\section{\label{sec:ooo}Ancilla-free, destructive continuous variable (CV) SWAP test}

In principle, the DV operator $V$ can be extended to a countably infinite dimensional Hilbert space $\mathcal{H}\otimes \mathcal{H}$ of two quantum harmonic oscillators (i.e., two CV registers). However, the computational basis states $\ket{j}$ correspond to Fock states, so the action of $V^{\dagger}$ on, e.g., a tensor product of CV coherent states, would produce optically non-classical, non-Gaussian states. In particular, $V^{\dagger}$ would not be implementable on a near-term photonic processor such as Xanadu's X8, which produces non-classical, Gaussian states \cite{arra}. In this section, we seek an ancilla-free CV SWAP test that requires only linear optical operations and photon number-resolving measurement.

The \texttt{SWAP} gate for continuous variables can be written as
\begin{equation}
\text{\texttt{SWAP}}=e^{i\pi \left( {a_{1}^{\dagger}-a_{2}^{\dagger}\over \sqrt{2}}\right)\left( {a_{1}-a_{2}\over \sqrt{2}}\right)}.
\label{eqn:swapg}
\end{equation}
As in the case of $\swp$ on a tensor product of isomorphic finite dimensional Hilbert spaces, (\ref{eqn:swapg}) is both unitary and self-adjoint. A generalized version of $\swp$ has been experimentally implemented for two-mode CV systems in a circuit quantum electrodynamics framework \cite{gao}. A proposal for implementing the ancilla-full CV SWAP test using a fault-tolerant ancilla prepared in a Kerr cat logical qubit appears in Ref.\cite{girvin}.

To prove (\ref{eqn:swapg}), one verifies that the unitary has the correct action on the canonical operators, $\text{\texttt{SWAP}}a^{\dagger}_{1(2)}\text{\texttt{SWAP}}^{\dagger}=a^{\dagger}_{2(1)}$. One can see that the \texttt{SWAP} gate is a linear optical unitary (i.e., its generator is quadratic in the creation and annihilation operators and commutes with the total photon number $\sum_{j=1,2}a^{\dagger}_{j}a_{j}$), and can therefore be written as a composition of 50:50 beamsplitters and local phase shifters according to the rectangular decomposition \cite{walmsley}. For example, defining the beamsplitter as in (\ref{eqn:cvst}), one gets
\begin{equation}
    \swp = e^{i{\pi\over 2}(a^{\dagger}_{1}a_{1}+a^{\dagger}_{2}a_{2})}U_{\text{BS}}({\pi\over 2},-{\pi\over 2}).
    \label{eqn:swgate}
\end{equation}
Note that $U_{\text{BS}}({\pi\over 4},0)\ket{n}_{1}\ket{m}_{2}$ is an eigenvector of $\swp$ with eigenvalue $(-1)^{n}$, and that these form a complete set of eigenvectors. 
Thus we can estimate the expectation value of $\swp$ using the circuit in Fig.~\ref{fig:aaa}d, in which only the photon count $n$ in the first register contributes to the overlap estimator.
Note that this is inherently different from the particular DV SWAP test algorithm  discussed in the previous section, where the eigenvalue of $\swp$ is determined by the measurement outcomes on both registers. 
One might try to reverse engineer an ancilla-free DV SWAP test algorithm for qudits based on the present CV algorithm. Alas, this is not possible because the beamsplitter unitary maps some states in the qudit subspace outside this subspace. This demonstrates that calculating some functions of qudit states might be much easier when embedded in a CV system, due to the fact that the state is occasionally allowed to leave the subspace associated with the qudit.

Further, \texttt{SWAP} in \eqref{eqn:swgate} commutes with the projections $Q_{2M}$ of subspaces spanned by $\lbrace \ket{n}_{A}\ket{m}_{B}: n+m\le 2M\rbrace$ and, as a consequence, also commutes with its complement $Q_{2M}^C=\mathbb{I}-Q_{2M}$. The set of states $\lbrace U_{\text{BS}}({\pi\over 4},0)\ket{n}_{A}\ket{m}_{B} : n+m\le 2M\rbrace$ is therefore an alternative orthonormal basis for the subspace $Q_{2M}\mathcal{H}\otimes \mathcal{H}$  associated with $Q_{2M}$. 
We define two operators that implement $\swp$ on the $Q_{2M}\mathcal{H}\otimes \mathcal{H}$ subspace and its complement via
\begin{align} 
\swp_{2M} &\equiv Q_{2M}\, \swp \, Q_{2M}\nonumber \\
&= \sum_{n+m\le 2M}\left[ (-1)^{n}U_{\text{BS}}\left( {\pi\over 4},0\right)  \ket{n}_{A}\ket{m}_{B}\bra{n}_{A}\bra{m}_{B}U_{\text{BS}}\left( {\pi\over 4},0\right)^{\dagger}\right]\nonumber \\
\swp_{2M}^C &\equiv Q_{2M}^C\, \swp \, Q_{2M}^C
\end{align}
which satisfy $\swp = \swp_{2M}+\swp_{2M}^C$.

Consider states $\rho$ and $\sigma$ such that $\text{supp }\rho\otimes\sigma$ is contained in $Q_{2M}\mathcal{H}\otimes \mathcal{H}$. Then
 \begin{align}
 \text{tr}\rho\sigma &= \text{tr}\swp  \rho\otimes \sigma \nonumber \\
 &=  \text{tr}\swp Q_{2M}\rho\otimes \sigma Q_{2M}\nonumber\\
 &= \text{tr} \swp_{2M} (\rho\otimes \sigma).
\label{eqn:cvsw}
\end{align}
Therefore, photon number-resolving detection with a finite experimental threshold on the total number of photons can be used to estimate the expectation of $\swp$ in $\rho\otimes \sigma$ granted all photon numbers above the threshold have zero probability. In an experiment, if the input states have local Fock cutoff $M'$, the local photon number-resolving detectors should have a threshold of at least $2M'$ photons, i.e., \eqref{eqn:cvsw} holds for $M\ge M'$. Thus $2M'$ is the minimal photon detector threshold that allows to apply the ancilla-free, CV SWAP test for input states with local Fock cutoff $M'$.

This fact provides an opportunity to concretely illustrate how the ancilla-free CV SWAP test takes into account amplitudes on all $2M'$ Fock basis states to estimate the overlap. Consider $M'=1$ and the task of estimating the overlap of single photon states $\ket{\psi}=\alpha_{1}\ket{0}+\beta_{1}\ket{1}$ and $\ket{\phi}=\alpha_{2}\ket{0}+\beta_{2}\ket{1}$ with our algorithm. The overlap is a homogeneous polynomial of order 4 in $\alpha_{1}$, $\overline{\alpha_{1}}$, $\alpha_{2}$, $\overline{\alpha_{2}}$, $\beta_{1}$, $\overline{\beta_{1}}$, $\beta_{2}$, $\overline{\beta_{2}}$, and this polynomial contains the term $\vert \beta_{1}\vert^{2}\vert\beta_{2}\vert^{2}$ whenever both $\beta_{1}$ and $\beta_{2}$ are nonzero. But writing out $U_{\text{BS}}(\pi/4,0)^{\dagger}\ket{\psi}\ket{\phi}$, one sees that there is no way to get this term without registering both the $\ket{0}\ket{2}$ and $\ket{2}\ket{0}$ photon count outcomes with +1 weight. Physically, these outcomes can be considered as ``Hong-Ou-Mandel'' contributions to the overlap. More generally, when $\rho$ and $\sigma$ have support on Fock states up to $\ket{M'}$, one needs to take $M\ge M'$ in the expression (\ref{eqn:cvsw}) so that all photon interferences arising from $U_{\text{BS}}({\pi\over 4},0)^{\dagger}$ contribute to the overlap estimate.

To rigorously extend the above argument to CV states $\rho\otimes \sigma$ with arbitrary support requires some analysis in separable Hilbert space. For instance, the photon number operator needs to have its domain defined, its spectral theorem should be stated, etc. We will forego the formalities, simply noting that $\lbrace \ket{n}\rbrace_{n=0}^{\infty}$ is a countable orthonormal basis for a CV mode $\ell^{2}(\mathbb{C})$ as discussed in Section \ref{sec:cvintro}. Consider pure CV states $\ket{\psi}$ and $\ket{\phi}$. Their respective amplitudes in the above orthonormal basis are square summable by Parseval identity, so $\lim_{M\rightarrow \infty}Q_{2M}\ket{\psi}_{A}\ket{\phi}_{B} = \ket{\psi}_{A}\ket{\phi}_{B}$. Define two sequences of  states $\xi_{2M}\equiv {1\over q_{2M}}Q_{2M}\ket{\psi}_{A}\ket{\phi}_{B}\bra{\psi}_{A}\bra{\phi}_{B}Q_{2M}$ and $\zeta_{2M}\equiv {1\over 1-q_{2M}}(\mathbb{I}-Q_{2M})\ket{\psi}_{A}\ket{\phi}_{B}\bra{\psi}_{A}\bra{\phi}_{B}(\mathbb{I}-Q_{2M})$, where $q_{2M}:=\bra{\psi}\bra{\phi}Q_{2M}\ket{\psi}\ket{\phi}$ allows to define the respective normalization factors. Then
\begin{align}
    \vert \langle \psi \vert\phi\rangle\vert^{2}&= \text{tr}\swp \ket{\psi}\bra{\psi}_{A}\otimes \ket{\phi}\bra{\phi}_{B}\nonumber \\
    &= \text{tr}(\swp_{2M}+\swp_{2M}^C) \ket{\psi}\bra{\psi}_{A}\otimes \ket{\phi}\bra{\phi}_{B}\nonumber \\
    &= q_{2M}\text{tr} \swp_{2M} \xi_{2M} + (1-q_{2M})\text{tr}\swp_{2M}^C \zeta_{2M}
    \label{eqn:full}
\end{align}
where the second line uses the fact that $(\mathbb{I}-Q_{2M})\swp Q_{2M} = (\mathbb{I}-Q_{2M})Q_{2M}\swp =0$.
Using (\ref{eqn:cvsw}), the result (\ref{eqn:full}) is extended by bilinearity (in the tensor factors $A$ and $B$) to normal states \cite{bratteli1} $\rho_{A}\otimes \sigma_{B}$ of two CV modes. The error due to Fock cutoff can be bounded as:
\begin{align}
    \big\vert \text{tr}\rho\sigma - \text{tr} \swp_{2M}(\rho\otimes \sigma)) \big\vert &= \vert \text{tr}\swp_{2M}^C (\rho\otimes \sigma) \vert \nonumber \\
        &= \vert \text{tr}\swp Q_{2M}^C(\rho\otimes \sigma)Q_{2M}^C \vert \nonumber \\
    &\le  \text{tr}Q_{2M}^C(\rho\otimes \sigma)Q_{2M}^C  \nonumber \\
        &\le 1-q_{2M}     \label{eqn:nnn} \\
        &\le 1 - q^\rho_M q^\sigma_M
        \label{eqn:bds}
    \end{align}
where $q^{\rho}_M=\text{tr}(\rho \sum_{n=0}^{M}\ket{n}\bra{n})$ and similarly for $q^{\sigma}_M$.
Again from the Parseval identity, it follows that the left hand side of (\ref{eqn:nnn}) can be made arbitrarily small by taking a large enough value of $M$. 
Thus the accuracy of the estimate obtained using a photon number-resolving measurement depends on $M$ and the precision depends on the number of shots $S$. 

Consider the task of estimating $\text{tr}(\rho \sigma)$ with some error $\epsilon$ using an experimental setup whose photon detectors on each register have a threshold given by $2M$. We perform the procedure shown in Fig.~\ref{fig:aaa}d and obtain two photon counts on each register: $n$ and $m$. 
We assign the value 0 to all realizations for which $n+m> 2M$. 
Note that for all values $n+m< 2M$ the photon detectors are not saturated. 
To these instances we assign the value $(-1)^{n}$ and compute the following estimator:
\begin{align}
    \reallywidehat{\text{tr}(\swp_{2M}\rho \otimes \sigma)} &= \frac{1}{S}\sum_{s=1}^S (-1)^{n(s)} \Theta[2M-n(s)-m(s)]
    \label{eqn:estimator2}
\end{align}
where $\Theta$ is the Heaviside step function with $\Theta(0)=1$.  Note that the Fock basis plays the role of the computational basis in the ancilla-free DV SWAP test of Section \ref{sec:rev}, although not all photon number counts are utilized in the estimator. Estimator (\ref{eqn:estimator2}) is the generalization of (\ref{eqn:estimator1}) with $K=1$ to the CV setting with the device cutoff taken into account. 
This is an unbiased estimator of the expectation value of $ \swp_{2M}$ operator.
Note that the expectation value of a quantum observable can be estimated with standard deviation $\delta$ by perfectly measuring the observable $O(1/\delta^2)$ number of times. 
The total error of the estimator $\vert \text{tr}\rho\sigma - \reallywidehat{\text{tr}(\swp_{2M}\rho \otimes \sigma)}\vert$ can be upper bounded by the sum of the systematic error $\vert \text{tr}\rho\sigma - \text{tr}\swp_{2M}(\rho\otimes \sigma)\vert$ and the statistical error $\vert \reallywidehat{\text{tr}(\swp_{2M}\rho \otimes \sigma)} - \text{tr}\swp_{2M}(\rho\otimes \sigma)\vert$.
In order to bound the total error by $\epsilon$ it suffices to choose $M$ such that $1-q_{2M} < \epsilon$ and run the algorihms with $S=O(1/\delta^2)$ shots where $0<\delta\le \epsilon-(1-q_{2M})$. The first condition puts limits on the quality of acceptable photon detectors whereas the second condition dictates the sampling complexity of the algorithm.

\begin{figure}[t]
\centering
\includegraphics[scale=0.6]{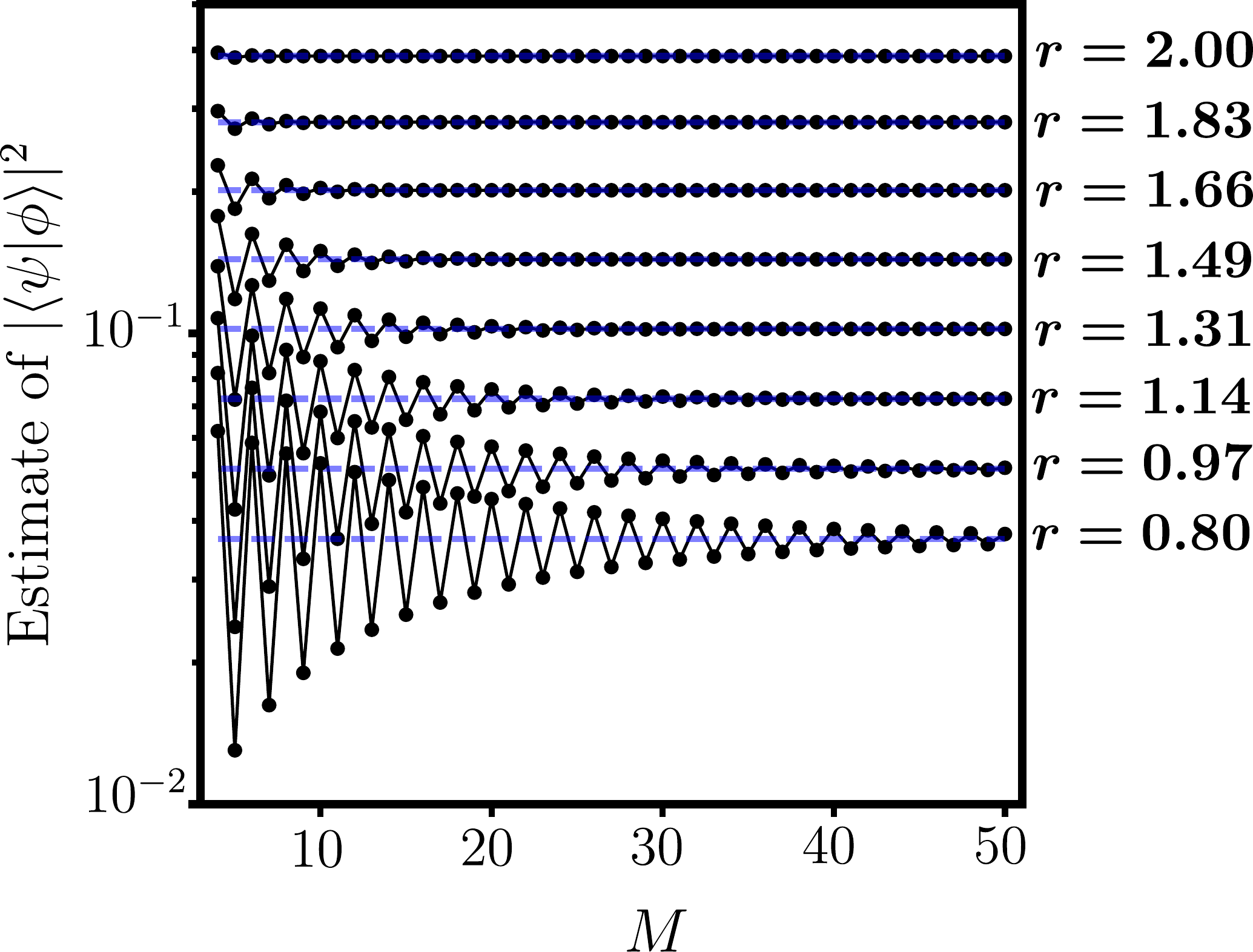}
	\caption{\label{fig:ttt}Blue dashed lines are ${1\over \cosh 2r}$ for eight values of $r\in [0.8,2]$. Black dots are 
	$\text{tr} \swp_{2M}(\rho\otimes \sigma)$ given by formula (\ref{eqn:sqp}) for the integer values $M=4,5,\ldots,50$ (black lines are guides to the eye).}
\end{figure}

It is useful to carry out a detailed error analysis for some example state pairs. First consider the unitary squeezing operator $S(r)=e^{{r\over 2}(a^{2}-a^{\dagger 2})}$ \cite{mandel}. With $\ket{\psi}_{A}\equiv S(r)\ket{0}_{A}$, $\ket{\phi}_{B}=S(-r)\ket{0}_{B}$, i.e., squeezed and anti-squeezed vacuum, one gets $\vert \langle \psi\vert\phi\rangle\vert^{2}={1\over \cosh 2r}$. Applying the circuit in Fig.~\ref{fig:aaa}d produces 
\begin{align}
&{}U_{BS}({\pi\over 4},0)^{\dagger}S(r)\ket{0}_{A}\otimes S(-r)\ket{0}_{B} = U_{TM}(r)\ket{0}_{A}\ket{0}_{B} \nonumber \\
&={1\over \cosh r}\sum_{n=0}^{\infty}(-\tanh r)^{n}\ket{n}_{A}\ket{n}_{B}
\end{align}
where $U_{\text{TM}}(r) =e^{r(ab-a^{\dagger}b^{\dagger})}$ is the two-mode squeezing operator. The finite Fock cutoff approximation to $\vert \langle \psi\vert\phi\rangle\vert^{2}$ is given by 
\begin{align}
    \text{tr} (\swp_{2M}\rho\otimes \sigma)  &=\begin{cases} {1+\tanh^{2(M+1)}r \over \cosh2r}& \text{ even } M\\
    {1-\tanh^{2(M+1)}r \over \cosh2r} & \text{ odd } M
    \end{cases}
\label{eqn:sqp}
\end{align}
with error upper bounded by (\ref{eqn:nnn})
\begin{align}
        1-q_{2M}&= 1-\Vert Q_{2M}S(r)\ket{0}_{A}\otimes S(-r)\ket{0}_{B}\Vert^{2}\nonumber \\
    &= \tanh^{2(M+1)}r.
    \label{eqn:tme}
\end{align}
Convergence of (\ref{eqn:sqp}) to $1/\cosh 2r$ is shown in Fig.~\ref{fig:ttt} for $r\in [0.1,2]$. 
From (\ref{eqn:tme}) and the bound in (\ref{eqn:nnn}), it follows that for large $r$, $M={e^{2r}\over 4}\ln (\epsilon^{-1}) -1$ is sufficient to get an error $\epsilon$ in the overlap. Physically, this is $M=O(E\ln \epsilon^{-1})$ where $E$ is the number of photons in the input state. 

Another example consists of computing $\vert \langle \alpha \vert \beta\rangle\vert^{2}$, where $\ket{\alpha}$ and $\ket{\beta}$ are any two isoenergetic coherent states of the quantum harmonic oscillator with energy $E$, i.e., $\vert \alpha\vert^{2}=\vert \beta \vert^{2}=E$. The total photon number random variable $n+m$ is Poisson distributed with mean $2E$. Therefore, an upper bound on $1-q_{2M}$ is obtained from an upper bound on the tail probability of the Poisson distribution. From a weak tail bound such as $1-q_{2M}\le 1-e^{-{E\over M}}$ ($M>E$), one can only infer that $M=O(E/\epsilon)$ is sufficient to obtain an error $\epsilon$ in the estimate of $\vert \langle \alpha \vert \beta\rangle\vert^{2}$. This scaling is worse than $\log \epsilon^{-1}$ that was obtained in the squeezed state example. However, the Chernoff bound gives that for $M>E$,
\begin{align}
    1-q_{2M}&\le \left( {eE\over M} \right)^{2M}e^{-2E} \nonumber\\
    &= (eE)^{2M(1-{\ln M\over 1+\ln E})}e^{-2E}.
    \label{eqn:uppupp}
\end{align}
One verifies numerically that for all $E$ and for $\epsilon \in (0,1)$ taking $M={13\over 10}E+\log \epsilon^{-1}$ results in (\ref{eqn:uppupp}) being less than $\epsilon$. 
It follows that for a fixed energy $E$, the smallest acceptable Fock cutoff for CV SWAP test for antipodal coherent states has a milder dependence on additive error $\epsilon$ compared to the CV SWAP test for squeezed and anti-squeezed states.

 To obtain the bound in (\ref{eqn:uppupp}) and derive an expression for $M$ that implies that the bound is less than $\epsilon$ requires knowledge of specific results on Poisson tail probabilities. For sufficiently large $E$, a general approach can be implemented that only utilizes the tail of the normal distribution. This approach is based on the fact that the looser bound (\ref{eqn:bds}) on the systematic error allows one to obtain an upper bound on the error from the local cumulative Fock distributions of $\rho$ and $\sigma$, respectively. For the case of antipodal coherent states, one first assumes $E$ large enough to justify the normal approximation to the Poisson distribution, viz., $\mathop{P}_{X\sim \text{Pois}(E)}(X\le M)\approx \mathop{P}_{X\sim \mathcal{N}(E,\sqrt{E})}(X\le M)$, then obtains the upper bound $1-\Phi\left( {M-E\over \sqrt{E}}\right)^{2}$ from (\ref{eqn:bds}). This upper bound is less than or equal to $\epsilon$ for
\begin{align}
    M&\ge E+\sqrt{E}\Phi^{-1}\left(\sqrt{1-\epsilon}\right)\nonumber \\
    &\ge E+\sqrt{\pi c E\over 8}\text{logit}(\sqrt{1-\epsilon})\nonumber \\
    &\sim  E+\sqrt{\pi c E\over 8}\ln(2\epsilon^{-1})
    \label{eqn:nbnbn}
\end{align}
for $\epsilon\rightarrow 0$, where $c<1$. In \eqref{eqn:nbnbn}, $\Phi^{-1}$ is the quantile function for the standard normal distribution (i.e., the probit function) and  $\text{logit}(x)\equiv \ln\left( {x\over 1-x}\right)$ on $(0,1)$.

\subsection{Experimental demonstration\label{sec:x8}}

To provide a proof-of-principle demonstration of the ancilla-free CV SWAP test, it would be ideal to implement an experiment showing convergence of the estimator (\ref{eqn:estimator2}) with different states and energy ranges. However, such an experiment is not presently possible using cloud-based CV processors such as Xanadu's X8 device \cite{arra}, which restricts the input state, decomposition of linear optical circuit, and photon number-resolving measurement cutoff. In particular, one cannot probe different energy ranges. Further, on the optical chip, a linear optical unitary is always precompiled to the rectangular decomposition, and there is no option to change the decomposition (the optical chip is not reconfigurable).  Specifications of the optical chip and photon number-resolving measurement for the X8 can be found in Ref.\cite{arra}; for the present purposes we note that the photodetectors have a quantum efficiency above 95\% and a local photon number cutoff between 5 and 7. 

Putting aside the limitations of the X8 device for doing a full experimental analysis, it is possible to implement the CV SWAP circuit in order to verify that it produces an approximation to the analytical result (recall that the multimode ancilla-free CV SWAP test simply requires applying the beamsplitter $U_{\text{BS}}({\pi\over 4},0)^{\dagger}$ to appropriate registers, followed by photon number-resolving measurement). The beamsplitter is implemented on the X8 in a noise-resistant way by specifying the rectangular decomposition (up a complex multiple of modulus 1)
\begin{equation}
    U_{\text{BS}}\left( {\pi\over 4},0\right)^{\dagger}=e^{i\pi a^{\dagger} a}U_{\text{MZ}}\left( {\pi\over 2},0\right)
\end{equation}
where $U_{\text{MZ}}(\phi_{1},\phi_{2})\equiv U_{\text{BS}}({\pi\over 4},{\pi\over 2})e^{i\phi_{1}a^{\dagger}a}U_{\text{BS}}({\pi\over 4},{\pi\over 2})e^{i\phi_{2}a^{\dagger}a}$ is the MZ (Mach-Zehnder) gate \cite{walmsley,volkcompile}. Note that we do not have access to the exact noise channels that affect the output of the device (i.e., the photon number-resolving measurement statistics).
The simplest nontrivial fidelity that can be estimated on the X8 is
\begin{equation}
    \big\vert \left( \vert 0 \rangle^{\otimes 2},\vert \text{TMSS}_{r}\rangle \right)\big\vert^{2}={1\over \cosh^{2}r}
\end{equation}
where $\ket{\text{TMSS}_{r}}={1\over\cosh r}\sum_{n=0}^{\infty}(-1)^{n}\tanh^{n}r\ket{n}\ket{n}$ is a two-mode squeezed state. According to the two-mode generalization of (\ref{eqn:estimator2}), an estimate of this overlap is obtained as
\begin{equation}
    {1\over S}\sum_{s=1}^{S}(-1)^{n_{A_{1}}(s)+n_{B_{1}}(s)}.
    \label{eqn:x8est}
\end{equation}

\begin{figure*}[t]
\centering
\includegraphics[scale=.9]{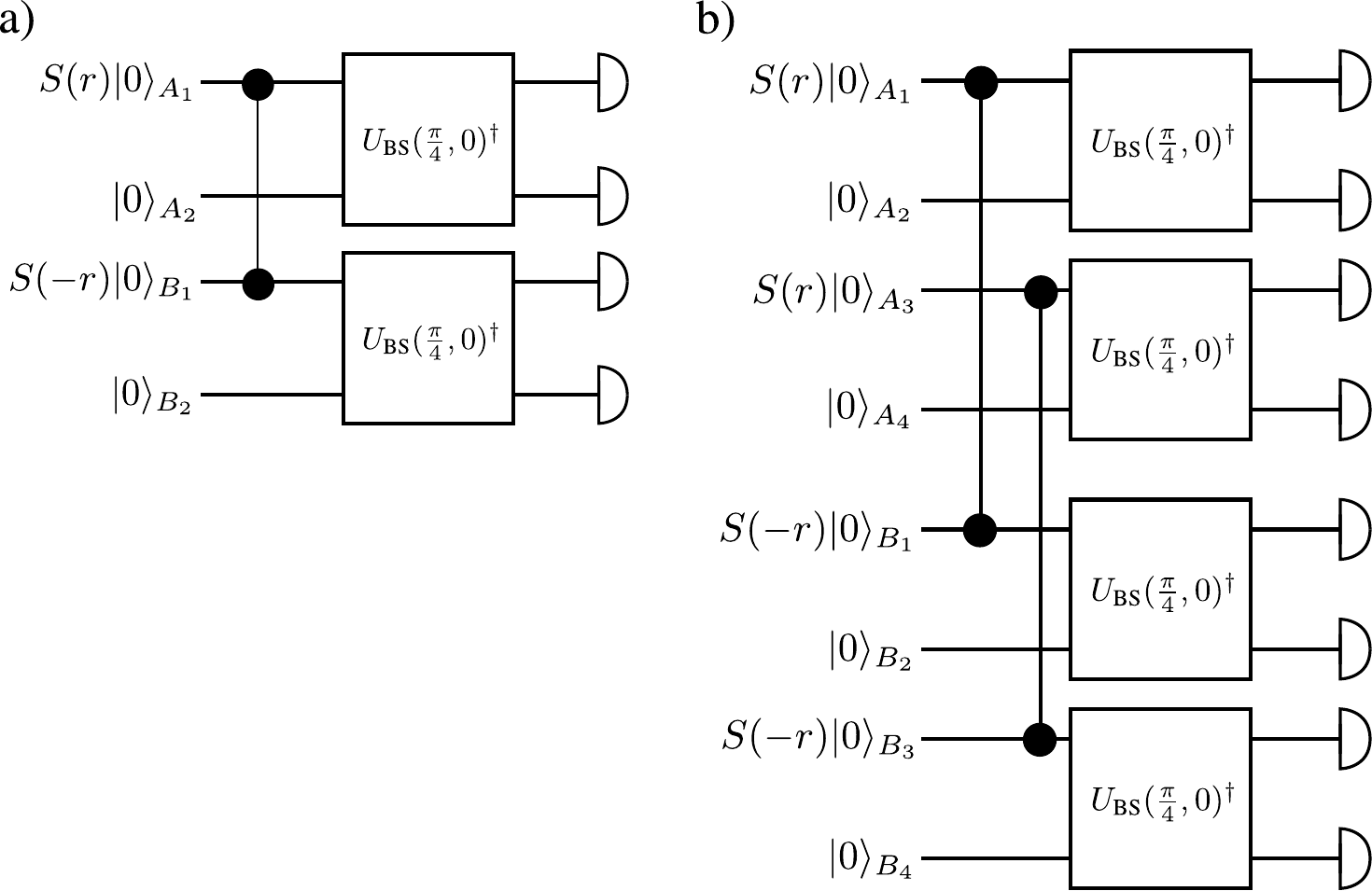}
	\caption{\label{fig:x8circ}Circuits for computing a) Eq.(\ref{eqn:x8est}) and b) Eq.(\ref{eqn:x8est2}) on the Xanadu's X8. $S(r)$ is the unitary quadrature squeezing operator and the squeezed states are coupled by 50:50 beamsplitter (dumbbell-shaped circuit elements) to form two-mode squeezed states. The squeezing factor $r\approx 1$ is fixed according to the device specifications. The two-mode gate denoted by the vertical dumbbell is $U_{\text{BS}}(-{\pi\over 4},0)$.}
\end{figure*}
Compared to (\ref{eqn:estimator2}), we are considering applying the ancilla-free CV SWAP circuit on two mode pairs, and we further omit the Heaviside step function because we use $r=1$, so the photon counts never exceed the photon detection threshold.
The circuit for computing (\ref{eqn:x8est}) is shown in Fig.~\ref{fig:x8circ}a. On the actual X8, the squeezing factor $r$ is constrained to $r\approx 1$. A numerical simulation of this circuit 
(10 runs, 2000 shots per run, local Fock cutoff 11) gave a mean 0.423 and standard deviation estimate 0.01, close to the analytical value ${1\over \cosh^{2}r} \approx 0.420$. Running the circuit of Fig.~\ref{fig:x8circ}a on the X8  produces a mean 0.4811 with standard deviation estimate 0.065 over 5 runs of $5\times 10^{4}$ shots per run. 

It is possible to utilize all 8 modes of the X8 processor to estimate in parallel the fidelity
\begin{equation}
    \big\vert \left( \vert 0\rangle^{\otimes 4}, \vert \text{TMSS}_{r}\rangle^{\otimes 2} \right)\big\vert^{2}={1\over \cosh^{4}r}
\end{equation}
from the estimator
\begin{equation}
    {1\over S}\sum_{s=1}^{S}(-1)^{n_{A_{1}}(s)+n_{B_{1}}(s)+n_{A_{3}}(s)+n_{B_{3}}(s)}.
    \label{eqn:x8est2}
\end{equation} The circuit for computing (\ref{eqn:x8est2}) is shown in Fig.~\ref{fig:x8circ}b. A numerical simulation of this circuit 
(10 runs, 500 shots per run, local Fock cutoff 10) gave a mean 0.176 and standard deviation estimate 0.03, close to the analytical value $1/\cosh^{4}(1)\approx 0.1764$. Running the circuit of Fig.~\ref{fig:x8circ}b on the X8 gives a large systematic difference from the analytical result, producing a mean 0.292 with standard deviation estimate 0.007 over 10 runs of $5\times 10^{4}$ shots per run. This systematic error could be due to the noise photons that occur even in modes initialized to vacuum, or from the cumulative effect of photon losses when many of the modes of the X8 are not initialized to vacuum.

The following section discusses applications of the ancilla-free CV SWAP test in the context of multimode CV quantum computing.

\section{Applications\label{sec:applic}}

 \subsection{Variational quantum compiling\label{sec:spec}}

The CV SWAP test provides an alternative method for computing cost functions that appear in the task of variational quantum compiling of CV circuits \cite{cvbpl,PRXQuantum.2.040327}. Such cost functions have the general form
\begin{equation}
    C_{T}(\theta)=1-{1\over K}\sum_{j=1}^{K}\vert \langle \psi_{j} \vert V(\theta)^{\dagger}U\otimes \mathbb{I}_{R}\vert \psi_{j}\rangle\vert^{2}
    \label{eqn:cvqaqc}
\end{equation}
where the training set $T\equiv \lbrace \ket{\psi_{j}} \rbrace_{j=1}^{K}$ is a set of states of a two mode CV system $\mathcal{H}_{A}\otimes\mathcal{H}_{R}$, and $\theta$ is shorthand for a set of continuous and discrete parameters that is optimized in order to minimize $C_{T}(\theta)$. Often, the states $\ket{\psi_{j}}$ are related by an energy-preserving symmetry operation. Parallel computation of the sum in (\ref{eqn:cvqaqc}) can be obtained by first preparing the $4K$-mode input state $\ket{\psi_{\text{in}}}\equiv \bigotimes_{j=1}^{K}\ket{\psi_{\text{in}}^{(j)}}$, where \begin{equation} \ket{\psi_{\text{in}}^{(j)}}\equiv U_{A_{j}}V(\theta)_{A_{j}'}\ket{\psi_{j}}_{A_{j}R_{j}}\ket{\psi_{j}}_{A'_{j}R'_{j}}\end{equation}
where $A_{j},A_{j}'\cong A$, $R_{j}, R_{j}'\cong R$. The circuit in Fig.~\ref{fig:aaa}d is then applied to all mode pairs $A_{j}A_{j}'$ and $R_{j}R_{j}'$  (the CV circuit architecture may require that beamsplitters be applied to adjacent modes, in which case the appropriate $\swp$ gates should be applied). This procedure works because the states 
\begin{equation}
    U_{\text{BS}}({\pi\over 4},0)_{AA'}U_{\text{BS}}({\pi\over 4},0)_{RR'} \ket{n}_{A}\ket{m}_{A'}\ket{n'}_{R}\ket{m'}_{R'}
\end{equation}
are eigenvectors of $\swp_{AA'}\swp_{RR'}$ with eigenvalue $(-1)^{n+n'}$. Assume for simplicity that for all $j$, $\ket{\psi^{(j)}_{\text{in}}}$ is contained in the subspace $Q_{2M_{j}}\mathcal{H}_{A_{j}}\otimes \mathcal{H}_{R_{j}}\otimes \mathcal{H}_{A_{j}'}\otimes \mathcal{H}_{R_{j}'}$ where $Q_{2M_{j}}$ projects to the subspace with orthonormal basis $\lbrace \ket{n}_{A_{j}}\ket{m}_{A_{j}'}\ket{n'}_{R_{j}}\ket{m'}_{R_{j}'}:n+m+n'+m' \le 2M_{j}\rbrace$. Under this assumption, the $j$-th term of the sum in (\ref{eqn:cvqaqc}) is given by 
\begin{align}
    \text{tr} 
Q_{2M_{j}}\swp_{A_{j}A_{j}'}\swp_{R_{j}R_{j}'}Q_{2M_{j}}\ket{\psi_{\text{in}}^{(j)}}\bra{\psi_{\text{in}}^{(j)}}  
\end{align} 
An estimator of this quantity can be constructed analogous to \eqref{eqn:estimator2} and, for general $\ket{\psi^{(j)}_{\text{in}}}$ without a finite Fock cutoff. 
An error analysis similar to previous section yields lower bounds on the sufficient photon count thresholds $M_{j}$ needed in order to get an additive error $\epsilon$.

\subsection{Two-copy test\label{sec:tct}}
\label{sec:twocopytest}

The CV SWAP test immediately gives a CV generalization of the two-copy test \cite{entspec}. 
The latter is based on the observation that for pure states $\ket{\Psi}$, the squared expectation value $|\bra{\Psi}U\ket{\Psi}|^2$ for some unitary operator $U$ is equal to the overlap between states $\ket{\Psi}$ and $U\ket{\Psi}$. Note that both real and imaginary parts of $\bra{\Psi}U\ket{\Psi}$ can be estimated using the Hadamard test which also works for mixed states. However, the two-copy test has some advantages. 
Unlike the Hadamard test which requires controlled-$U$, the two-copy test only uses $U$. In general, controlled versions of unitaries require more complicated (larger depth) quantum circuits, which for noisy devices can be detrimental. Moreover, a controlled unitary cannot be enacted if one only has black-box access to the unitary \cite{Araujo2014}.  The two-copy test is useful whenever one does not know how to measure in the eigenbasis of $U$, but $U$ is easy to implement. 

We now describe the two-copy test in the CV setting. For a state $\ket{\Psi}$ on $n$ mode CV register $C$ and a CV unitary $U$ acting on $C'$, one computes $\vert \langle \Psi \vert U\vert\Psi\rangle\vert^{2}$ by preparing $\ket{\Psi}_{C}\otimes U\ket{\Psi}_{C'}$ with $C\cong C'$, and applying the ancilla-free, $2$-mode CV SWAP test (the circuit is Fig.~\ref{fig:aaa}d on each mode pair $C_{j}C_{j}'$, $j=1,\ldots, n$). The full circuit has depth 2.

\begin{figure*}[t]
\centering
\includegraphics[scale=1.9]{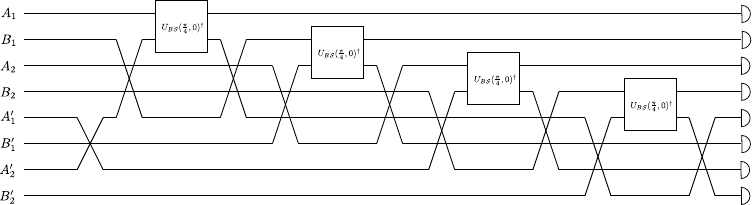}
	\caption{\label{fig:renyi}Circuit for estimation of $(\text{tr}\rho_{A}^{2})^{2}$ via two-copy test on a photonic processor that only allows beamsplitting between nearest-neighbor modes. The $\swp_{A_{1}'A_{2}'}$ gate is $\texttt{PERM}_{A'}$, which implements the cyclic permutation. The remaining $\swp$ gates allow to implement the beamsplitters according to the assumed nearest-neighbor architecture.}
\end{figure*}

As an example, the CV two-copy test with $U$ a cyclic permutation of many CV modes can be employed for computing $(\text{tr}\rho^{n})^{2}$, $n\ge2$. This was also a primary application of the DV two-copy test \cite{entspec}.  Specifically, one takes  $C$ to be a CV register $C=AB=A_{1}B_{1}\cdots A_{n}B_{n}$ and $U$ to be the CV cyclic permutation $\texttt{PERM}_{A}\equiv \prod_{j=1}^{n}\swp_{A_{j}A_{j+1}}$  (where $n+1$ is taken modulo $n$) on the $n$-mode register $A$ consisting of isomorphic modes $A_{1},\ldots, A_{n}$ ($\rho$ can be considered as a state on $A_{j}$ for any $j$).  
We let $\ket{\Psi}_{C}\equiv \bigotimes_{j=1}^{n}\ket{\psi}_{A_{j}B_{j}}$, $j=1,\dots, n$, and $\ket{\psi}$ is a purification of $\rho$. We prepare a second copy $\ket{\Psi}_{C'}$ and the state $\ket{\Psi}_{C}\otimes \ket{\Psi}_{C'}$ is then subjected to $\texttt{PERM}_{A'}$, i.e., a cyclic permutation of the subsystem $A'$ in the second copy.
Finally, the ancilla-free CV SWAP test of Section \ref{sec:ooo} is carried out on the state $\ket{\Psi}_{C}\otimes \texttt{PERM}_{A'}\ket{\Psi}_{C'}$.  This algorithm works because analogous to \eqref{eqn:swpswp}
\begin{equation}
 \langle \Psi_{C} \vert \texttt{PERM}_{A} \vert \Psi_{C} \rangle=\text{tr}\rho^{n}\, .
\label{eqn:cycperm}
\end{equation}
One can see that this CV two-copy test produces an estimate of $(\text{tr}\rho^{n})^2$.
For $n=2$,  the full circuit for computing (\ref{eqn:cycperm}) from input $\ket{\psi}_{A_{1}B_{1}}\ket{\psi}_{A_{2}B_{2}}\ket{\psi}_{A_{1}'B_{1}'}\ket{\psi}_{A_{1}'B_{1}'}$ is as shown in Fig.~\ref{fig:renyi}. More generally, for a pure CV state $\ket{\psi}_{C_{1}\cdots C_{n}}$ on $n$ CV modes, the CV two-copy test allows to  compute the quantities $(\text{tr}\rho_{\Sigma}^{n})^{2}$, where $\Sigma \subset \lbrace C_{1},\ldots, C_{n}\rbrace$. 

It should be noted that $\texttt{PERM}_{A'}$ does not need to be implemented as a gate as it amounts to a reindexing of the registers. This is the advantage of the two-copy test over the Hadamard test, where $\texttt{PERM}_{A'}$ is controlled on an ancilla and has to be implemented on the device. 
Nevertheless, one may still have to implement some $\swp$ gates for the two-copy test in devices with limited connectivity. We note that a constant-depth version of the Hadamard test has been proposed that makes use of constant-depth preparation of ancilla Greenberger-Horne-Zeilinger (GHZ) states and constant-depth implementation of of a multiply controlled $\texttt{PERM}$ \cite{wilde}.

\subsection{Extension of SWAP test and entanglement spectroscopy\label{sec:ext}}

We now discuss a novel CV algorithm for calculation of $\text{tr}(\rho^{L})$ ($L\ge 2$) that, to the knowledge of the authors, does not have a strict analogue for DV systems. It makes use of the fact that a short depth implementation of a measurement in the orthonormal basis of eigenvectors of $\texttt{PERM}$ is possible in CV systems. To see this, note that similar to the construction of eigenvectors of the CV $\swp$ gate in Section \ref{sec:ooo}, one can construct a linear optical unitary $U$ such that $U\ket{n_{0}}_{0}\otimes \cdots \otimes \ket{n_{L-1}}_{L-1}$ are eigenvectors of $\texttt{PERM}$ (here we are using a $L$ mode CV system). One just uses the fact that $\texttt{PERM}$ acts on the vector $(a_0^{\dagger},\ldots,a_{L-1}^{\dagger})$ via a circulant matrix, so  defining $U$ such that
\begin{equation}
    Ua_{j}^{\dagger}U^{\dagger}=b_{j}^{\dagger}\equiv {1\over \sqrt{L}}\sum_{\ell=0}^{L-1}e^{2\pi i \ell j \over L}a_{\ell}^{\dagger}
\end{equation}
one finds that the states
\begin{equation} \prod_{j=0}^{L-1}{b_{j}^{\dagger n_{j}}\over \sqrt{n_{j}!}}\ket{\text{VAC}}
\end{equation}
are eigenvectors of $\texttt{PERM}$ with eigenvalues $\prod_{j=0}^{L-1}e^{2\pi i j n_{j}\over L}$. The relation between $b_{j}^{\dagger}$ and the $a_{\ell}^{\dagger}$ is a discrete Fourier transform on $\lbrace 0 ,\ldots, L-1\rbrace$.  The unitary $U$ can be compiled on a linear optical circuit with local phase shifters and nearest-neighbor 50:50 beamsplitters by using the  rectangular form \cite{walmsley}. In particular, there is a depth $L$ circuit of beamsplitters and phase shifters that compiles $U$. Such a linear optical operation has also been proposed for implementing an order-$M$ SWAP test without ancillas for single-photon states, which consumes a state of the form $\ket{\phi}\otimes \ket{\psi}^{\otimes M-1}$ and allows to estimate $\vert \langle \phi\vert\psi \rangle\vert^{2}$ by postprocessing photon number-resolving detection patterns  \cite{PhysRevA.98.062318}.

Applying $\text{Ad}_{U^{\dagger}}$ to input state $\rho = \rho^{(0)}\otimes \cdots \otimes \rho^{(L-1)}$ allows one to obtain an estimate of $\text{tr}\texttt{PERM} \rho = \text{tr}\prod_{\ell=0}^{L-1}\rho^{(\ell)}$ from the estimator
\begin{equation}
   \reallywidehat{ \text{tr}\texttt{PERM} \rho}={1\over S}\sum_{s=1}^{S}\prod_{j=0}^{L-1}e^{2\pi i j n_{j}(s)\over L}
\end{equation}
where $n_j{(s)}$ is the photon count on the $j$-th register at the $s$-th run of the experiment. 
We refer to this extension of the ancilla-free SWAP test as PERM test. 
The case of $\rho^{(i)}=\rho$ for $i=1,\dots,L$ corresponds to  estimating $\text{tr}\rho^{L}$ which can be used in determining the entanglement spectrum when $\rho$ is a reduced state of a larger quantum system \cite{entspec}.

To the best of our knowledge, a short-depth (linear in $L$) quantum circuit for diagonalizing the cyclic permutation operator \texttt{PERM} is not known for qubit systems. As a result, the two-copy test has been used for estimating $(\text{tr}\rho_{A}^{L})^{2}$ from $2L$ copies of a purification $\ket{\psi}_{AB}$ of $\rho_{A}$ \cite{entspec,entspec2}. That algorithm can also be used in the CV case using results of Section \ref{sec:twocopytest}. It has the advantage of having constant depth (independent of $L$) but it is inferior to the PERM test in other ways. First, the PERM test does not require purification of the quantum states. Second, PERM test requires half the number of modes the two-copy test does. And third, the sampling complexity of two-copy test is larger than that of PERM test because the former estimates the square of the desired quantity which is less than one \cite{entspec}. Finally, we also note that the functions $\text{tr}\rho_{A}^{L}$ are used to upper bound the entanglement entropy of a subregion $\Sigma$ of a lattice spin system, and can be numerically estimated in these systems by using quantum Monte Carlo to sample the $\swp_{\Sigma}$ operator in such systems \cite{PhysRevLett.104.157201}. In fact, path integral Monte Carlo methods for sampling $\swp_{\Sigma}$ have been applied to estimate $\text{tr}\rho_{\Sigma}^{2}$, $\Sigma \subset \mathbb{R}^{3}$, for massive particles in the continuum \cite{herdman}.

\subsection{DV-CV hybrid SWAP test\label{sec:dvcv}}

Combining the DV SWAP test in Section \ref{sec:rev} and the CV SWAP test in Section \ref{sec:ooo}, one obtains an ancilla-free, destructive SWAP test that can be applied in hybrid quantum systems, such as cavity QED. In more detail, and specializing the DV system to be a qubit system, let $A$ be a qubit register and $B$ a CV register, and consider the task of estimating $\vert \langle \phi\vert\psi\rangle\vert^{2}$ where $\ket{\psi}_{AB}$ and $\ket{\phi}_{AB}$ are pure states of the hybrid system. The ancilla-free, destructive hybrid SWAP test proceeds by: 1. preparation of the state $\ket{\psi}_{AB}\ket{\phi}_{A'B'}$, 2. application of the circuit in Fig.~\ref{fig:aaa}b to the modes $AA'$ and the circuit in Fig.~\ref{fig:aaa}d to the modes $BB'$, 3. computational basis measurement of $AA'$ and photon number-resolving measurement of $BB'$. Combining (\ref{eqn:estimator1}) and (\ref{eqn:estimator2}), the estimator of $\text{tr}\rho\sigma$ given by
\begin{align}
    &{}\reallywidehat{\text{tr}\rho\sigma} =  \frac{1}{S} \sum_{s=1}^S (-1)^{i_{A}(s)j_{A'}(s)+n_{B}(s)}\Theta[2M-n_{B}(s)-m_{B'}(s)]
    \label{eqn:hyb}
\end{align}
where $(i_{A},j_{A'})$ is a two-bit word associated with an outcome of a Bell measurement. Eq.(\ref{eqn:hyb}) has arbitrarily small error for sufficiently large $M$. The proof of this statement follows from noting that if $\ket{\psi}\ket{\phi}$ has  support in $(\mathbb{C}^{2})_{A}\otimes (\mathbb{C}^{2})_{A'}\otimes (Q_{2M})_{BB'}(\mathcal{H}_{B}\otimes \mathcal{H}_{B'})$, then
\begin{align}
    \vert \langle \psi\vert \phi\rangle\vert^{2}&= \text{tr}\left[ \vphantom{\sum_{j=0}^{\infty}} \swp_{AA'}\swp_{BB'}  (\mathbb{I}_{AA'}\otimes (Q_{2M})_{BB'})\right. \nonumber \\
    &{} \left. \ket{\psi}_{AB}\ket{\phi}_{A'B'}\bra{\psi}_{AB}\bra{\phi}_{A'B'} (\mathbb{I}_{AA'}\otimes (Q_{2M})_{BB'})\vphantom{\sum_{j=0}^{\infty}}\right]
\end{align}
holds exactly, where the $\swp$ gates are the qubit and CV versions, respectively. One can then follow the analysis of (\ref{eqn:full}) and (\ref{eqn:nnn}) to get an upper bound on the error of the estimate (\ref{eqn:hyb}) when $\ket{\psi}\ket{\phi}$ have arbitrary support. Extensions of this hybrid algorithm to the applications in Sections \ref{sec:spec} and  \ref{sec:tct} is a matter of bookkeeping of all DV and CV quantum registers.

\section{Discussion\label{sec:discussion}}

The destructive, ancilla-free CV SWAP test presented in this work is envisioned as an equality testing subalgorithm \cite{mon} for CV states that can be embedded in larger CV quantum algorithms. Our discussions of applications such as variational quantum compiling and implementation in hybrid DV-CV systems indicate potential implementations in near-term quantum hardware. Indeed, we were able to carry out a demonstration on the X8 device, thereby showing the feasibility of the proposed CV SWAP test for few CV modes on a noisy photonic chip.

We note that a SWAP test for high-dimensional motional states of trapped ions has recently been implemented \cite{swapexp}, further suggesting the importance of the SWAP test beyond the qubit context. Advantages of the present proposal include: no ancilla degrees of freedom or controlled gates, fully linear optical depth 1 circuit, and photon number-resolving measurement that can be achieved using transition-edge sensors. These requirements are presently available on photonic chips and free-space optical setups which manipulate low-energy continuous variable states. Technologies that extend these systems to a wider range of frequency modes and energies will allow for practical implementation of the present algorithm in more quantum optical settings.

\acknowledgements

The authors thank M. Wilde and P. Coles for helpful discussions and the anonymous referees for useful suggestions. 
This work was supported by the U.S. Department of Energy through the Los Alamos National Laboratory. Los Alamos National Laboratory is operated by Triad National Security, LLC, for the National Nuclear Security Administration of U.S. Department of Energy (Contract No. 89233218CNA000001).
The authors were supported by the Laboratory Directed Research and Development (LDRD) program of Los Alamos National Laboratory (LANL).
This material is based upon work supported by the U.S. Department of Energy, Office of Science, National Quantum Information Science Research Centers.

\bibliographystyle{quantum}
\bibliography{refs}

\begin{thebibliography}{10}

\bibitem{PhysRevLett.97.110501}
Nicolas~C. Menicucci, Peter van Loock, Mile Gu, Christian Weedbrook, Timothy~C.
  Ralph, and Michael~A. Nielsen.
\newblock ``Universal quantum computation with continuous-variable cluster
  states''.
\newblock \href{https://dx.doi.org/10.1103/PhysRevLett.97.110501}{Phys. Rev.
  Lett. {\bf 97}, 110501}~(2006).

\bibitem{PhysRevLett.112.120504}
Nicolas~C. Menicucci.
\newblock ``Fault-tolerant measurement-based quantum computing with
  continuous-variable cluster states''.
\newblock \href{https://dx.doi.org/10.1103/PhysRevLett.112.120504}{Phys. Rev.
  Lett. {\bf 112}, 120504}~(2014).

\bibitem{PhysRevX.8.021054}
Kosuke Fukui, Akihisa Tomita, Atsushi Okamoto, and Keisuke Fujii.
\newblock ``High-threshold fault-tolerant quantum computation with analog
  quantum error correction''.
\newblock \href{https://dx.doi.org/10.1103/PhysRevX.8.021054}{Phys. Rev. X {\bf
  8}, 021054}~(2018).

\bibitem{PhysRevA.101.012316}
Kyungjoo Noh and Christopher Chamberland.
\newblock ``{Fault-tolerant bosonic quantum error correction with the
  surface--Gottesman-Kitaev-Preskill code}''.
\newblock \href{https://dx.doi.org/10.1103/PhysRevA.101.012316}{Phys. Rev. A
  {\bf 101}, 012316}~(2020).

\bibitem{Bourassa2021blueprintscalable}
J.~Eli Bourassa, Rafael~N. Alexander, Michael Vasmer, Ashlesha Patil, Ilan
  Tzitrin, Takaya Matsuura, Daiqin Su, Ben~Q. Baragiola, Saikat Guha, Guillaume
  Dauphinais, Krishna~K. Sabapathy, Nicolas~C. Menicucci, and Ish Dhand.
\newblock ``Blueprint for a {S}calable {P}hotonic {F}ault-{T}olerant {Q}uantum
  {C}omputer''.
\newblock \href{https://dx.doi.org/10.22331/q-2021-02-04-392}{{Quantum} {\bf
  5}, 392}~(2021).

\bibitem{dj1}
A.~K. Pati and S.~L. Braunstein.
\newblock ``{Deutsch-Jozsa Algorithm for Continuous Variables}''~(2002).
\newblock
  \href{http://arxiv.org/abs/quant-ph/0207108}{arXiv:quant-ph/0207108}.

\bibitem{dj2}
R.~C. Wagner and V.~M. Kendon.
\newblock ``{The continuous-variable Deutsch–Jozsa algorithm using realistic
  quantum systems}''.
\newblock \href{https://dx.doi.org/10.1088/1751-8113/45/24/244015}{J. Phys. A:
  Math. Theor. {\bf 45}, 244015}~(2012).

\bibitem{grov}
A.~K. Pati, S.~L. Braunstein, and S.~Lloyd.
\newblock ``{Quantum searching with continuous variables}''~(2000).
\newblock
  \href{http://arxiv.org/abs/quant-ph/0002082}{arXiv:quant-ph/0002082}.

\bibitem{PhysRevLett.87.167902}
Harry Buhrman, Richard Cleve, John Watrous, and Ronald de~Wolf.
\newblock ``Quantum fingerprinting''.
\newblock \href{https://dx.doi.org/10.1103/PhysRevLett.87.167902}{Phys. Rev.
  Lett. {\bf 87}, 167902}~(2001).

\bibitem{PhysRevA.87.052330}
Juan~Carlos Garcia-Escartin and Pedro Chamorro-Posada.
\newblock ``{SWAP test and Hong-Ou-Mandel effect are equivalent}''.
\newblock \href{https://dx.doi.org/10.1103/PhysRevA.87.052330}{Phys. Rev. A
  {\bf 87}, 052330}~(2013).

\bibitem{cincio}
L.~Cincio, Y.~Suba\c{s}{\i}, A.~T. Sornborger, and P.~J. Coles.
\newblock ``Learning the quantum algorithm for state overlap''.
\newblock \href{https://dx.doi.org/10.1088/1367-2630/aae94a}{New J. Phys. {\bf
  20}, 113022}~(2018).

\bibitem{PhysRevResearch.3.033251}
Kok~Chuan Tan and Tyler Volkoff.
\newblock ``{Variational quantum algorithms to estimate rank, quantum
  entropies, fidelity, and Fisher information via purity minimization}''.
\newblock \href{https://dx.doi.org/10.1103/PhysRevResearch.3.033251}{Phys. Rev.
  Research {\bf 3}, 033251}~(2021).

\bibitem{PhysRevB.96.195136}
Sonika Johri, Damian~S. Steiger, and Matthias Troyer.
\newblock ``Entanglement spectroscopy on a quantum computer''.
\newblock \href{https://dx.doi.org/10.1103/PhysRevB.96.195136}{Phys. Rev. B
  {\bf 96}, 195136}~(2017).

\bibitem{entspec}
Y.~Suba\c{s}{\i}, L.~Cincio, and P.~J. Coles.
\newblock ``Entanglement spectroscopy with a depth-two quantum circuit''.
\newblock \href{https://dx.doi.org/10.1088/1751-8121/aaf54d}{J. Phys. A: Math.
  Theor. {\bf 52}, 044001}~(2019).

\bibitem{entspec2}
J.~Yirka and Y.~Suba\c{s}{\i}.
\newblock ``Qubit-efficient entanglement spectroscopy using qubit resets''.
\newblock \href{https://dx.doi.org/10.22331/q-2021-09-02-535}{Quantum {\bf 5},
  535}~(2021).

\bibitem{brun}
T~Brun.
\newblock ``Measuring polynomial functions of states''.
\newblock \href{https://dx.doi.org/10.26421/QIC4.5-6}{Quantum Information
  Processing {\bf 4}, 401}~(2004).

\bibitem{PhysRevX.9.031013}
Jordan Cotler, Soonwon Choi, Alexander Lukin, Hrant Gharibyan, Tarun Grover,
  M.~Eric Tai, Matthew Rispoli, Robert Schittko, Philipp~M. Preiss, Adam~M.
  Kaufman, Markus Greiner, Hannes Pichler, and Patrick Hayden.
\newblock ``Quantum virtual cooling''.
\newblock \href{https://dx.doi.org/10.1103/PhysRevX.9.031013}{Phys. Rev. X {\bf
  9}, 031013}~(2019).

\bibitem{PhysRevX.11.041036}
William~J. Huggins, Sam McArdle, Thomas~E. O'Brien, Joonho Lee, Nicholas~C.
  Rubin, Sergio Boixo, K.~Birgitta Whaley, Ryan Babbush, and Jarrod~R. McClean.
\newblock ``Virtual distillation for quantum error mitigation''.
\newblock \href{https://dx.doi.org/10.1103/PhysRevX.11.041036}{Phys. Rev. X
  {\bf 11}, 041036}~(2021).

\bibitem{holmes2021nonlinear}
Zo{\"e} Holmes, Nolan Coble, Andrew~T Sornborger, and Yi{\u{g}}it
  Suba{\c{s}}{\i}.
\newblock ``On nonlinear transformations in quantum computation''~(2021).
\newblock  \href{http://arxiv.org/abs/2112.12307}{arXiv:2112.12307}.

\bibitem{bratteli2}
O.~Bratteli and D.~W. Robinson.
\newblock ``Operator algebras and quantum statistical mechanics 2''.
\newblock Springer, Berlin. ~(1997).

\bibitem{holevo}
A.~S. Holevo.
\newblock ``Quantum systems, channels, information''.
\newblock de Gruyter, Berlin/Boston. ~(2012).

\bibitem{PhysRevLett.104.157201}
Matthew~B. Hastings, Iv\'an Gonz\'alez, Ann~B. Kallin, and Roger~G. Melko.
\newblock ``{Measuring Renyi Entanglement Entropy in Quantum Monte Carlo
  Simulations}''.
\newblock \href{https://dx.doi.org/10.1103/PhysRevLett.104.157201}{Phys. Rev.
  Lett. {\bf 104}, 157201}~(2010).

\bibitem{durrett}
R.~Durrett.
\newblock ``Probability, 4th ed.''.
\newblock Cambridge University Press. ~(2010).

\bibitem{arra}
J.~M. Arrazola, V.~Bergholm, K.~Bradler, T.~R. Bromley, M.~J. Collins,
  I.~Dhand, A.~Fumagalli, T.~Gerrits, A.~Goussev, L.~G. Helt, J.~Hundal,
  T.~Isacsson, R.~B. Israel, J.~Izaac, S.~Jahangiri, R.~Janik, N.~Killoran,
  S.~P. Kumar, J.~Lavoie, A.~E. Lita, D.~H. Mahler, M.~Menotti, B.~Morrison,
  S.~W. Nam, L.~Neuhaus, H.~Y. Qi, N.~Quesada, A.~Repingon, K.~K. Sabapathy,
  M.~Schuld, D.~Su, J.~Swinarton, A.~Szava, K.~Tan, P.~Tan, V.~D. Vaidya,
  Z.~Vernon, Z.~Zabaneh, and Y.~Zhang.
\newblock ``Quantum circuits with many photons on a programmable nanophotonic
  chip''.
\newblock \href{https://dx.doi.org/10.1038/s41586-021-03202-1}{Nature {\bf
  591}, 54}~(2021).

\bibitem{gao}
Yvonne~Y. Gao, Brian~J. Lester, Kevin~S. Chou, Luigi Frunzio, Michel~H.
  Devoret, Liang Jiang, S.~M. Girvin, and Robert~J. Schoelkopf.
\newblock ``Entanglement of bosonic modes through an engineered exchange
  interaction''.
\newblock \href{https://dx.doi.org/10.1038/s41586-019-0970-4}{Nature {\bf 566},
  509}~(2019).

\bibitem{girvin}
I.~Pietikainen, O.~Cernatik, S.~Puri, R.~Filip, and S.~M. Girvin.
\newblock ``Controlled beam splitter gate transparent to dominant ancilla
  errors''.
\newblock \href{https://dx.doi.org/10.1088/2058-9565/ac760a}{Quantum Sci.
  Technol. {\bf 7}, 035025}~(2022).

\bibitem{walmsley}
William~R. Clements, Peter~C. Humphreys, Benjamin~J. Metcalf, W.~Steven
  Kolthammer, and Ian~A. Walmsley.
\newblock ``Optimal design for universal multiport interferometers''.
\newblock \href{https://dx.doi.org/10.1364/OPTICA.3.001460}{Optica {\bf 3},
  1460--1465}~(2016).

\bibitem{bratteli1}
O.~Bratteli and D.~W. Robinson.
\newblock ``Operator algebras and quantum statistical mechanics 1''.
\newblock Springer, Berlin. ~(1987).

\bibitem{mandel}
L.~Mandel and E.~Wolf.
\newblock ``Optical coherence and quantum optics''.
\newblock Cambridge University Press. ~(1995).

\bibitem{volkcompile}
T.~J. Volkoff.
\newblock ``{Strategies for variational quantum compiling of a zero-phase
  beamsplitter on the Xanadu X8 processor}''~(2022).
\newblock  \href{http://arxiv.org/abs/2202.01161}{arXiv:2202.01161}.

\bibitem{cvbpl}
T.~J. Volkoff.
\newblock ``Efficient trainability of linear optical modules in quantum optical
  neural networks''.
\newblock \href{https://dx.doi.org/10.1007/s10946-021-09958-1}{J. Russ. Laser
  Res. {\bf 42}, 250}~(2021).

\bibitem{PRXQuantum.2.040327}
Tyler Volkoff, Zo\"e Holmes, and Andrew Sornborger.
\newblock ``Universal compiling and (no-)free-lunch theorems for
  continuous-variable quantum learning''.
\newblock \href{https://dx.doi.org/10.1103/PRXQuantum.2.040327}{PRX Quantum
  {\bf 2}, 040327}~(2021).

\bibitem{Araujo2014}
Mateus Ara{\'{u}}jo, Adrien Feix, Fabio Costa, and {\v{C}}aslav Brukner.
\newblock ``Quantum circuits cannot control unknown operations''.
\newblock \href{https://dx.doi.org/10.1088/1367-2630/16/9/093026}{New Journal
  of Physics {\bf 16}, 093026}~(2014).

\bibitem{wilde}
Y.~Quek, E.~Kaur, and M.~M. Wilde.
\newblock ``{Multivariate trace estimation in constant quantum depth}''~(2022).
\newblock  \href{http://arxiv.org/abs/2206.15405}{arXiv:2206.15405}.

\bibitem{PhysRevA.98.062318}
Ulysse Chabaud, Eleni Diamanti, Damian Markham, Elham Kashefi, and Antoine
  Joux.
\newblock ``Optimal quantum-programmable projective measurement with linear
  optics''.
\newblock \href{https://dx.doi.org/10.1103/PhysRevA.98.062318}{Phys. Rev. A
  {\bf 98}, 062318}~(2018).

\bibitem{herdman}
C.~M. Herdman, P.-N. Roy, R.~G. Melko, and A.~Del~Maestro.
\newblock ``{Entanglement area law in superfluid $^{4}$He}''.
\newblock \href{https://dx.doi.org/10.1038/NPHYS4075}{Nat. Phys. {\bf 13},
  556}~(2017).

\bibitem{mon}
Ashley Montanaro and Ronald~{de} Wolf.
\newblock ``A survey of quantum property testing''.
\newblock \href{https://dx.doi.org/10.4086/toc.gs.2016.007}{Pages 1--81}.
\newblock Number~7 in Graduate Surveys. Theory of Computing Library. ~(2016).

\bibitem{swapexp}
C.-H. Nguyen, K.-W. Tseng, G.~Gleb~Maslennikov, H.~C.~J. Gan, and
  D.~Matsukevich.
\newblock ``{Experimental SWAP test of infinite dimensional quantum
  states}''~(2021).
\newblock  \href{http://arxiv.org/abs/2103.10219}{arXiv:2103.10219}.

\end{thebibliography}

\onecolumn\newpage
\appendix

\section{$\swp$ eigenvectors from qudit Bell states\label{sec:app1}}

In this appendix, we show that eigenvectors of $\swp$ have superposition rank at most 2 with respect to the qudit Bell states, which may prove relevant to a proposal for a short-depth circuit for an ancilla-free qudit SWAP test. An orthonormal basis consisting of eigenvectors of $\swp$ can be produced from qudit Bell states by implementing rotations in the two-dimensional subspaces spanned by any pair of qudit Bell states that have equal $z$. This follows again from the fact that $\text{\texttt{SWAP}}=P_{+}-P_{-}$ and $\text{\texttt{SWAP}}\ket{\Phi_{z,x}}=e^{-2\pi i xz /d}\ket{\Phi_{z,-x}}$. To see this, just note that $P_{\pm}={\mathbb{I}\pm\swp \over 2}$ so, for example in the case of odd $d$, the states
\begin{align}
    {\ket{\Phi_{z,x}} \pm \swp \ket{\Phi_{z,x}} \over \sqrt{2}} = {\ket{\Phi_{z,x}} \pm e^{-{2\pi ixz\over d}} \ket{\Phi_{z,-x}} \over \sqrt{2}}
    \label{eqn:www}
\end{align}
for $x\neq 0$ are normalized eigenvectors of $\swp$ with eigenvalue $\pm 1$ (for even $d$, $x={d\over 2}$ would not give a normalized state since ${d\over 2}=-{d\over 2}\mod d$). In full generality, one can define $\ket{\Omega_{z,x}^{\pm}}$, $x\notin \lbrace 0,{d\over 2}\rbrace$, to be the states in (\ref{eqn:www}) and note that for even $d$, an orthonormal set of $\swp$ eigenvectors is given by
\begin{align}
    &{}\lbrace \ket{\Phi_{z,0} }\rbrace_{z=0}^{d-1} \sqcup \lbrace \ket{\Phi_{2z,{d\over 2}}}\rbrace_{z=0}^{{d-2\over 2}} \sqcup \lbrace \ket{\Omega_{z,x}^{+}}\rbrace_{\substack{z=0,\ldots,d-1 \\ x=1,\ldots {d\over 2}-1}}\nonumber \\
    &\sqcup \lbrace \ket{\Omega_{z,x}^{-}}\rbrace_{\substack{z=0,\ldots,d-1 \\ x=1,\ldots {d\over 2}-1}} \sqcup \lbrace \ket{\Phi_{2z+1,{d\over 2}}}\rbrace_{z=0,\ldots,{d\over 2}-1}
    \label{eqn:swp1}
\end{align}
whereas for odd $d$ an orthonormal set of $\swp$ eigenvectors is given by
\begin{align}
    &{}\lbrace \ket{\Phi_{z,0} }\rbrace_{z=0}^{d-1}  \sqcup \lbrace \ket{\Omega_{z,x}^{+}}\rbrace_{\substack{z=0,\ldots,d-1 \\ x=1,\ldots {d-1\over 2}}}\sqcup  \ket{\Omega_{z,x}^{-}}\rbrace_{\substack{z=0,\ldots,d-1 \\ x=1,\ldots {d-1\over 2}}} .
    \label{eqn:swp2}
\end{align}

Therefore, given $d$, if $W$ is a circuit that generates the states in (\ref{eqn:swp1}) or (\ref{eqn:swp2}) from the qudit computational basis, then $W^{\dagger}$ can replace $V^{\dagger}$ in the qudit ancilla-free destructive SWAP test circuit. 


\end{document}